\documentclass[12pt]{aastex631}
\usepackage{hhline}
\usepackage{lineno}
\usepackage{tabularx}
\usepackage[]{hyperref}
\newcommand\setrow[1]{\gdef\rowmac{#1}#1\ignorespaces}
\newcommand\clearrow{\global\let\rowmac\relax}
\clearrow

\begin{document}

\title{Supernovae Origin for the Low-Latitude-Intermediate-Velocity Arch and the North-Celestial-Pole Loop}

\author[0000-0001-7908-6940]{J.T. Schmelz}
\affiliation{USRA, 425 3rd Street SW, Suite 950, Washington, D.C., joan.t.schmelz@gmail.com}

\author[0000-0002-6160-1040]{G.L. Verschuur}
\affiliation{verschuur@aol.com}

\author[0000-0003-3833-2513]{A. Escorza}
\affiliation{ European Southern Observatory, Alonso de C\'{o}rdova 3107, Vitacura, Santiago, Chile}

\author[0000-0002-1883-4578]{A. Jorissen}
\affiliation{Institut d’Astronomie et d’Astrophysique, Universit\'{e} Libre de Bruxelles (ULB), Bruxelles, Belgium}

\keywords{ISM: atoms - ISM: clouds - Stars: binaries - Stars: supernovae}

\begin{abstract}
Supernova explosions attributed to the unseen companion in several binary systems identified by the Third Gaia Data Release (Gaia DR3) may be responsible for a number of well-known and well-studied features in the radio sky including the Low-Latitude-Intermediate-Velocity Arch and the North-Celestial-Pole Loop. Slices from the Longitude-Latitude-Velocity data cube of the $\lambda$-21-cm galactic neutral hydrogen $HI4PI$ survey \citep{BenBekhti2016} show multiple signatures of an expanding shell. The source of this expansion, which includes the Low-Latitude-Intermediate-Velocity Arch on the approaching side, may be the neutron star candidate in the Gaia DR3 1093757200530267520 binary. If we make the simplifying assumptions that the expansion of the cavity is uniform and spherically symmetric, then the explosion took place about 700,000 years ago. The momentum is in reasonable agreement with recent model estimates for a supernova this old. The HI on the receding side of this cavity is interacting with the gas approaching us on the near side of a second cavity. The North-Celestial-Pole Loop appears to be located at the intersection of these two expanding features. The neutron star candidate in the Gaia DR3 1144019690966028928 binary may be (in part) responsible for this cavity. Explosions from other candidates may account for the observed elongation along the line of sight of this second cavity. We can use the primary star in these binaries to anchor the distances to the Low-Latitude-Intermediate-Velocity Arch and North-Celestial-Pole Loop, which are $\sim$167 and $\sim$220 pc, respectively.
\end{abstract}

\section{Introduction}

High-velocity clouds (HVCs) are concentrations of hydrogen gas found all over the sky, sometimes in large complexes. They are defined by their anomalous velocities, which are inconsistent with the regular rotation of the Milky Way Galaxy. Please see \cite{Verschuur1975} for an initial review and \cite{vanWoerden2004} for subsequent developments. The origins of HVCs are a mystery, mainly because their distances are unknown. Most models place them in the Galactic Halo, kiloparsecs from the Sun (see, e.g., \citealt{vanWoerden-Wakker2004}). At these vast distances, some of the complexes would contain millions of solar masses and contribute significantly to star formation and Galactic evolution. Some promising models – like old supernova explosions – are rejected simply because they cannot generate enough energy to propel this much mass to the observed high velocities \citep{Oort1966}. But what if HVCs, or at least some of them, were much closer? 

\cite{Schmelz2022} found that the HVC known as MI may be the result of a supernova that took place about 100,000 years ago at a distance of 163 pc. Low-velocity HI data show a clear cavity centered on the spatial coordinates of MI, ($l$, $b$) = (165$^{\circ}$, 65.$^{\circ}5$). The M1 Cavity is also visible in 100 $\mu$m dust data from IRAS. The invisible companion of the yellow giant star, 56 Ursae Majoris, may be the remains of the supernova \citep{Escorza2023} that evacuated the M1 Cavity and propelled MI outward at 120 km s$^{-1}$. The mass and energy of MI are easily in line with what is expected from a supernova, and the diffuse X-rays seen by ROSAT \citep{Herbstmeier1995} provide evidence of a hot cavity.

\cite{Verschuur2023} showed that many of the individual features of HVC Complex M, including MI, MIIa, and MIIb, are components of a long, arched filament that extends from about ($l$, $b$) = (105$^{\circ}$, 53$^{\circ}$) to ($l$, $b$) = (196$^{\circ}$, 55$^{\circ}$). They used velocity maps, Gaussian analysis, and high-energy data to show that the MI cloud and the arched filament are physically interacting. They then used the known distance to MI, 150 pc as reported by \cite{Schmelz2022}, to bootstrap the distance to Complex M and estimate the mass of about 120 M$_\odot$.

\cite{Schmelz2023} found the origin event that accelerated the neutral atomic hydrogen of Complex M. An associated cavity centered at ($l$, $b$) $\sim$ (150$^{\circ}$, 50$^{\circ}$) extends about $\pm$33$^{\circ}$ in Galactic latitude and longitude. Using the known distance to Complex M \citep{Verschuur2023} and assuming that the Complex M Cavity is spherical, they found a distance of 307 pc to the explosive source of the Complex M Cavity, which has a radius of 166 pc and an expansion velocity 40 km s$^{-1}$. The energy of the expanding Complex M Cavity is about 3.0 ${\times}$ 10$^{50}$ ergs, well within the range of a single supernova that took place about four million years ago. As the blast propagated outwards, it swept up interstellar gas and carved out the Local Chimney \citep{Welsh1999, Lallement2003}, a low-density extension of the Local Bubble \citep{Zucker2022} that reaches all the way into the galactic halo. 

With these results in mind, we searched for other examples of anomalous velocity gas that might be explained by a supernova explosion. One promising example appears to be a region, ($l$, $b$) $\sim$ (110-180$^{\circ}$, 20-50$^{\circ}$), that includes/incorporates/interacts with several well-known and well-studied features in the radio sky including the Low-Latitude-Intermediate-Velocity (LLIV) Arch \citep{Kuntz1996}, the North Celestial Pole (NCP) Loop \citep{Heiles1989, Meyerdierks1991}, two high-latitude cirrus complexes \citep{Barriault2010}, and several high-latitude molecular clouds \citep{Magnani1985}. High Velocity Cloud (HVC) Complex A is also in this direction, but at much greater negative velocities \citep{Muller1963}. Please see \cite{Marchal2023} for a comprehensive set of references related to this area. 

In this paper, we explore the association of the HI gas dynamics with a possible supernova explosion. Although there may be as many as 10$^9$ neutron stars in the Milky Way Galaxy, only $\sim$10$^3$ have been observed directly as either pulsars or accreting X-ray sources. Although the target area contains multiple pulsars, all are either too distant, too old, or both \citep{Manchester2005}. 2005). Since there are no accreting X-ray binaries in the target area \citep{Liu2006, Liu2007}, and detections of non-radio-emitting isolated neutron stars like Geminga are rare, we have elected to search for neutron-star candidates that may be part of binary systems.

\section{Analysis}

\subsection{Radio Longitude-Latitude Maps}

The radio data used here are from the $\lambda$-21-cm galactic neutral atomic hydrogen $HI4PI$ survey \citep{BenBekhti2016}, which combines the northern hemisphere data from the Effelsberg-Bonn Survey ($EBHIS$; \citealt{Winkel2016}) and the southern hemisphere data from the third revision of the Galactic All-Sky Survey ($GASS$; \citealt{McClure-Griffiths2009}). $HI4PI$ has an angular resolution of $\Theta_{FWHM}\ =\ $ 16.\arcmin2, a sensitivity of $\sigma\ =\ $ 43 mK, and full spatial sampling, 5\arcmin\ in both galactic longitude ($l$) and latitude ($b$). 

Fig. 1 shows a series of maps of the HI brightness temperature at different velocities that include features that have traditionally been identified as the LLIV Arch \citep{Kuntz1996}. Panels (a) through (f) at -70, -66, -62, -58, -54, and -50 km s$^{-1}$, respectively, show different velocity slices of the approaching shell. The red dashed circle highlights the approximate boundary of the overall structure. As expected in almost any map of interstellar neutral hydrogen gas, there is a lot of complex seemingly interconnected structure here, which may be just a line-of-sight coincidence or may be related to the subject of our study. It is our challenge to determine which. 

The remaining panels of Fig. 1 show the low-velocity gas structure. Many of these images are dominated by the NCP Loop, which is located on the upper-right rim of the cavity and appears to be morphologically similar to an H II region blister \citep{Whitworth1979} or a prominence in the solar atmosphere. Fig. 1g at -2 km s$^{-1}$ shows the NCP Loop as well as a circular structure with multiple extending threads in the lower-left quadrant around ($l$, $b$) $\sim$ (160$^{\circ}$, 30$^{\circ}$). Fig. 1h at $+$2 km s$^{-1}$ shows the complex cavity structure and a more-defined image of the NCP Loop. Fig. 1i at $+$6 km s$^{-1}$ shows a circular structure with multiple extending threads in the upper-right quadrant, similar to the fainter structure seen in Fig. 1g, but this one is well-known and well-studied \citep{Barriault2010}. Fig. 1j at $+$10 km s$^{-1}$ shows the first hints of the oval cavity (blue dash) traditionally associated with the NCP Loop and several high-latitude molecular clouds \citep{Magnani1985, Pound1997}. Fig. 1k $+$14 km s$^{-1}$ shows the lower arc at ($l$, $b$) = (143$^{\circ}$, 25$^{\circ}$) that helps define the oval cavity. Fig. 1l at $+$18 km s$^{-1}$ shows higher-velocity traces of the oval cavity. To avoid confusion, we refer to the cavity outlined in red as the Circular Cavity and the one outlined in blue as the Oval Cavity throughout this paper.

\subsection{Radio Position-Velocity Maps}

The HI structures identified in the ($l$, $b$) maps of Fig. 1 can be clearly recognized in position – velocity maps. To cover the area described above, we used the $HI4PI$ data cube to produce 27 one-degree-wide latitude-velocity ($b$, $v$) maps every 2$^{\circ}$ from $l$ = 175$^{\circ}$ to 123$^{\circ}$ and 31 one-degree-wide longitude-velocity ($l$, $v$) maps every 1$^{\circ}$ from $b$ = 45$^{\circ}$ to 31$^{\circ}$. A selection from each set is shown in Fig. 2 and Fig. 3, which reveal a lot of complicated structure, including (1) a ridge of emission around $-$55 km s$^{-1}$ that forms part of the LLIV Arch; (2) the existence of the HI shell; (3) the NCP Loop; and (4) components of HVC Complex A.

Fig. 2 shows a series of ($b$, $v$) maps at different longitudes chosen to highlight the features discussed in Fig. 1. Fig. 2a at $l$ = 169$^{\circ}$, near the high-longitude boundary of the shell, shows twin ridges of positive- and near zero-velocity gas from the bottom of the plot to about $b$ = 35-38$^{\circ}$. The positive-velocity ridge may represent the receding component of an expanding shell. Fig. 2b at $l$ = 157$^{\circ}$ shows the HI emission ridge associated with the approaching side of the shell at a velocity around -55 km s$^{-1}$. This feature has a velocity gradient pattern (curvature), a signature of HI expanding away from the source and toward the observer. This structure connects to low-velocity gas at 28$^{\circ}$ and 44$^{\circ}$. The low-velocity emission around 0 km s$^{-1}$ appears relatively undisturbed. Fig. 2c at $l$ = 147$^{\circ}$ shows that the structure of the approaching shell at $l$ = 147$^{\circ}$ is extremely clumpy. The disturbance of the low-velocity gas in the latitude range 35 to 45$^{\circ}$ is gas on the far side of the star expanding away from the source of the explosion. Fig. 2d at $l$ = 143$^{\circ}$ cuts across the center of the shell outlined in Fig. 1. The distortion of the low-velocity gas is dramatic; the curvature is a signature of HI expanding away from the observer. On the near (negative velocity) side, the shell exhibits great irregularity where it may be interacting with ambient HI. (Note: The long, horizontal contour at 40.$^{\circ}$9 is the galaxy M 81.) In Fig. 2e at $l$ = 139$^{\circ}$, the near side of the shell shows knots and clumps. A connection to low-velocity gas is apparent at about 33$^{\circ}$ and 46$^{\circ}$. Despite localized irregularities, the shell-like pattern can be recognized in these data. Fig. 2f shows that the shell structure at $l$ = 133$^{\circ}$ around -55 km s$^{-1}$ is still evident, extending from 32$^{\circ}$ to 44$^{\circ}$. The bright structure at positive velocities up to $+$10 km s$^{-1}$ reveals the presence of the NCP Loop.

Fig. 3 shows a series of ($l$, $v$) maps at different latitudes chosen to highlight the features discussed in Fig. 1 and Fig. 2. Fig. 3a at $b$ = 39$^{\circ}$ shows the curvature of the HI emission around -50 to -60 km s$^{-1}$, a signature of the structure on the near side of the expanding shell. There are bridges to low-velocity gas at around 175$^{\circ}$ and 125$^{\circ}$. The low-velocity gas shows a clear distortion to positive velocities. Fig. 3b at $b$ = 37$^{\circ}$ shows that the emission from the shell is again well-defined, bridging to low-velocity gas at around $l$ = 130 and 170$^{\circ}$. The bright feature at +5 km s$^{-1}$ is part of the NCP Loop. This map also shows the distortion of the low-velocity peak to positive velocities from $l$ = 120$^{\circ}$ to 145$^{\circ}$. The low-velocity gas has a wavelike velocity gradient between the location of the emission from the NCP Loop area around $l$ = 120$^{\circ}$ to the edge of the map at $l$ = 180$^{\circ}$. Fig. 3c at $b$ = 35$^{\circ}$ shows the continuity of the HI around -55 km s$^{-1}$ linking with low-velocity gas at $l$ = 130$^{\circ}$ and 170$^{\circ}$. Emission from the NCP Loop is still prominent at around $l$ = 125$^{\circ}$.

Also visible on the left of Fig. 2 and Fig. 3 at velocities between -100 and -200 km s$^{-1}$ are components of HVC Complex A. Traditional absorption-line studies place Complex A at a distance of 10 kpc in the Galactic Halo \citep{vanWoerden-Wakker2004}. However, recent HI surveys reveal pervasive filamentary structure \citep{McClure-Griffiths2009, Peek2011, Winkel2016, Martin2015}. \cite{Mebold1990, Ryans1997} caution that the assumption of homogeneity over the entire radio beam required for the absorption-line analysis may be spurious, and a non-detection cannot provide a lower limit to the distance. As a result, Complex A could be much closer.

\subsection{Binary Target Selection}
\label{Sect:selection}

In order to find a faint compact object that could be the remnant of a supernova explosion, we searched for astrometric and spectroscopic binaries that could host neutron-star companions. First, we queried all systems flagged as non-single stars (NSS) in the Third Gaia Data Release (DR3; \citealt{GaiaDR3-NSS}) within 
the partially overlapping regions with radius $15^{\circ}$ centered on $l = 145^{\circ}, b = 34^{\circ}$   (corresponding to the Circular Cavity), and with radius $10^\circ$ centered on $l = 133^{\circ}, b = 32^{\circ}$  (corresponding to the Oval Cavity).
This choice was based on a visual inspection of the $HI4PI$ maps in Fig. 1l, while bearing in mind that important uncertainties such as the kick velocity and the precise age of the cavities prevent us from accurately identifying binaries that were close to the center of each cavity at the time of the explosion.

Within the Circular Cavity, \cite{GaiaDR3-NSS} identified 2109 spectroscopic binaries with one visible component, 2013 binaries with an astrometric orbital solution, and 788 systems with combined astrometric and spectroscopic solutions. The equivalent numbers for the Oval Cavity are 900, 1185, and 388.

We also queried the mass estimations and/or mass limits published as part of Gaia DR3 (in table \texttt{gaiadr3.binary\_masses}). These masses were computed using the orbital parameters and photometric information. To identify systems with possible neutron-star companions, when masses are available, we kept systems with $M_1<M_2$ and $M_2\ge1.1$~M$_{\odot}$ (see \citealt{Doroshenko2022} and \citealt{Chamel2013} for a discussion of the minimum -- 0.77~M$_\odot$ -- and maximum -- 2.5~M$_\odot$ -- neutron-star masses, respectively). In the present analysis, we will however consider 1.3~M$_{\odot}$ as the typical neutron-star mass. It should be noted that the use of the Gaia DR3 mass estimates in our selection process introduces a bias against systems hosting a giant primary, since masses for giant stars are lacking in the Gaia DR3 table \texttt{gaiadr3.binary\_masses}. More specifically, for spectroscopic binary systems with one available spectrum (SB1), the selection was based on the spectroscopic mass function $f(M_1,M_2) = \sin^3 i\; M_2^3/(M_1+M_2)^2$ being larger than 0.3~M$_\odot$. As shown in Table~\ref{Tab:fM}, this criterion will mostly select systems with low-mass primaries ($M_1 < 1.3$~M$_{\odot}$) and companions in the neutron-star mass range ($M_2\ge1.3$~M$_\odot$), although it is not exempt of false positives when neither $M_1$ nor the orbital inclination $i$ are known. This criterion should therefore not be given too much weight.   

For astrometric systems, a quantity similar to the spectroscopic mass function is the astrometric mass ratio function (AMRF) following equation 9 in \cite{GaiaDR3-NSS}:
\begin{equation}
    AMRF = \frac{a_0}{\varpi}\;M_1^{-1/3}\;P^{-2/3},
\end{equation}
where $a_0$ is the angular semi-major axis of the photocentric orbit, $\varpi$ is the parallax, $M_1$ is the mass of the primary star (in M$_\odot$), and $P$ is the orbital period of the binary (in yr).
Figure~33 of \cite{GaiaDR3-NSS} makes it clear that a compact low-luminosity companion is the only possible solution when $AMRF > 0.65$, independently of $M_1$. If $M_1$ is known, a more stringent criterion may be used. For systems with primaries of low masses ($0.5 \le M_1 \le 1$~M$_\odot$), compact low-luminosity companions are expected when $AMRF$ is larger than 
0.5 to 0.65 (however, those compact companions have masses $M_2$ more compatible with white dwarfs than neutron stars; see the right panel of Fig. 33 of \citealt{GaiaDR3-NSS}). For $M_1$ in the range 1 - 1.5~M$_\odot$, the $AMRF$ threshold decreases linearly from 0.65 to about 0.6, with $M_2$ falling among the typical values for neutron stars. Systems with $AMRF$ above this threshold are labelled $``$AMRF Class III$"$ by \cite{GaiaDR3-NSS}. 
As discussed by \cite{vandenHeuvel2020}, a difficulty in finding (non-accreting, i.e. $dormant$) neutron-star companions in binary systems on the basis of masses alone results from the possibility that the companion could consist of two main-sequence stars. 
The mass - luminosity relation ($L \propto M^{3.5}$) indicates that this pair of equal-mass main-sequence stars has a luminosity about 6 times smaller than that of a single star with their combined mass. 
AMRF Class III systems are precisely those for which such a configuration may be excluded due to constraints from the photometry (see \citealt{GaiaDR3-NSS} for details), and therefore, should be considered as the best candidates for hosting a neutron-star companion. 

\subsection{Neutron-star Candidates}

Tables~2 -- 4 
list all candidate systems that could host a neutron-star companion, according to the criteria described in Sect.~\ref{Sect:selection}. Table~2 summarizes their positional properties: 
equatorial and Galactic coordinates, parallax (either from the Gaia DR3 NSS processing in the case of Astrometric and AstroSpectroSB1 systems, or from the Gaia DR3 main table for SB1), distance (from a simple inversion of the parallax; no Bayesian estimate is necessary given that the relative parallax uncertainty is smaller than 5\%), and height $|z| = d \; \sin b$ above the Galactic plane.

Table~3 lists the binary properties related to our selection criteria: NSS type, masses of the components ($M_1$ and $M_2$ when available or the lower limit $M_{2,\mathrm{lower}}$ for SB1; for some astrometric binaries, only the range
$M_{2,\mathrm{lower}} - M_{2,\mathrm {upper}}$ is available), spectroscopic mass function $f(M_1,M_2)$, astrometric mass ratio function $AMRF$, center-of-mass velocity $V_{\rm cm}$ (only for SB1 and AstroSpectroSB1 systems; for astrometric systems, the radial velocity listed is from the Gaia main table), and space velocity
\begin{equation}\label{Vspace}
V_{\rm space} = (4.74^2 (\mu_{\alpha*}^2+\mu_{\delta}^2)/\varpi^2+V_{\rm cm}^2)^{1/2}.
\end{equation}
For $V_{\rm space}$ as for $\varpi$, the proper motions entering the computation of $V_{\rm space}$ are part of the NSS solution in the case of astrometric systems, but are taken from the 
Gaia main table for SB1 systems.
Period and eccentricity are listed next, and finally the angular on-sky distances from the neutron-star candidate to the center of the cavities (the adopted centers are $l = 145^\circ, b = 34^\circ$ and $l = 133^\circ, b = 32^\circ$ for the Circular and Oval Cavities, respectively), now and 100,000 yrs ago ($t_{SN}$), adopted as minimum time elapsed since the supernova explosion, given the absence of any visible emission. 

Table~4 lists the stellar parameters of the primary stars, including their Gaia $G$-band mean magnitude, their (non-dereddened) absolute magnitude $M_G$ in the Gaia band, their $B_p - R_p$ colour, their effective temperatures ($T_{\rm eff}$), their surface gravities ($\log g$) and their global metallicity $[M/H]$, according to Gaia DR3 $``$astrophysical parameters$"$
(preference has been given to the parameters derived from the Gaia spectrum whenever available
(labelled \texttt{gspspec} in the Gaia tables), photometrically-based values were used otherwise
(labelled \texttt{gspphot}). For comparison, Tables 2 -- 4 also list the same parameters for the two dormant black holes in binary systems discovered by Gaia DR3 (Gaia BH1 and BH2; \citealt{ElBadry2023-MSBH, ElBadry2023-RGBH}).


The candidates may be classified into five different categories, according to the decreasing likelihood that the candidate is truly a neutron star:
\begin{itemize}
    \item (i) The AMRF Class III systems, as defined by \citet[][also Sect.~\ref{Sect:selection}]{GaiaDR3-NSS}, are astrometric systems where $M_1$ and $M_2$ are both known, $M_1 < M_2$, $M_2 \ge 1.25$~M$_\odot$, and moreover, with photometry excluding the possibility that the companion is a pair of twin main-sequence stars. Such cases are \#13, 15, 18, and 28. System \#13 is especially interesting in that it lies close to the center of the Oval Cavity and its inferred mass is $1.39 \pm 0.12$~M$_\odot$. 
    
    Several among the Astrometric+SB1 and AstroSpectroSB1\footnote{Contrary to  AstroSpectroSB1 which were processed at once as astrometric and spectroscopic binaries by the NSS DR3 consortium, Astrometric+SB1 systems were processed separately as astrometric and SB1 systems.} (\#22, 24, 25, 32, 33, 34) may be added to this first category most likely hosting a neutron star, since the companion is several tenths of M$_\odot$ more massive than the primary despite not contributing substantially to the system light (since the system appears as SB1 rather than SB2). Given its central location in the Circular Cavity and its companion's inferred mass of $1.34 \pm 0.32$~M$_\odot$, System \#25 appears to be the best candidate for the progenitor of the supernova responsible for the formation of that cavity. 
    \item (ii) This second category is similar to the first, except for the fact that $M_1$ and $M_2$ are much closer together than for category (i) systems. Therefore their AMRF value is either below the AMRF threshold discussed above, or too close above it to tag them as AMRF Class III systems with confidence. Thus, despite the fact that the mass $M_2$ is compatible with a neutron star, there is a risk that the companion be a pair of (twin) main-sequence stars. Systems belonging to this category are \#1, 7, 11, 14, 27, and 29.
    \item (iii) This category contains the two entries \#10 and 31, which despite belonging to the NSS class AstroSpectroSB1 and Astrometric+SB1 respectively, have only $f(M)$ available but no individual masses from table \texttt{ gaiadr3.binary\_masses}, because the primary is a (sub)giant star with $\log g = 2.1$ and 2.4 respectively (Table~4). These systems would require a specific analysis to derive their masses as done by \citet{Escorza2023}. 
    \item (iv) The remaining astrometric binaries (\#2, 3, 4, 5, 8, 9, 12, 16, 17, 19, 20) have just lower and upper limits on their companion masses, and as such, are less well constrained than those from the previous classes. 
    \item (v) Finally, SB1 binaries (\#6, 21, 23, 26, 30) with no binary motion detected by astrometry are the least constrained, with only their mass functions $f(M)$ available, possibly with a lower limit on $M_2$ when $M_1$ may be estimated from astrophysical considerations. 
\end{itemize}

A major challenge to fully understanding these systems results from the fact that the current primary is always a low-mass star, while the former primary must have been a massive star ($>$~8~M$_{\odot}$) if it produced a supernova. For such a system to remain bound after the explosion, less than half of the initial mass may have been ejected \citep[e.g.,][]{Postnov2014}. To circumvent this problem, we need to invoke an asymmetric explosion, imparting a kick to both the exploding star, which strongly modifies the orbital properties like period and eccentricity \citep{Postnov2014, Escorza2023}, as well as the center-of-mass of the system \citep{Brandt1995, Kalogera1996}. It is worth mentioning that the two black hole binaries discovered so far using Gaia DR3 data also have low-mass primaries as well as orbital parameters outside the theoretically expected ranges \citep{Chakrabarti2022,ElBadry2023-RGBH,ElBadry2023-MSBH}. Their properties (Tables 2 -- 4) are quite different from previously known black hole binaries, and explaining their formation as well as their currently observed properties like periods, eccentricities, space velocities, etc. seems challenging as well.

To evaluate whether a natal kick could have been imparted to the neutron star at birth, with consequences on the system space velocity, Table~3 therefore lists both the systemic center-of-mass velocity ($V_{\rm cm}$ when available from the SB1 orbital elements, otherwise the RV from the Gaia DR3 main table is given) and the space velocity $V_{\rm space}$ (see Eq.~\ref{Vspace}). Special attention must be paid to systems with $V_{\rm space} > 80$~km/s, which could be runaway systems because of the supernova explosion \citep{Brandt1995, Kalogera1996, Fortin2022}. Interestingly, System \#13 (the AMRF Class III system located very close to the center of the Oval Cavity) has a large $V_{\rm space}$ of 80 km/s, which is unlikely to be attributable to its Galactic rotation given its relative proximity (488 pc). The same holds true for System \#25, at the center of the Circular Cavity, albeit with a slightly lower value of $V_{\rm space}$ (74.4 km/s).

We note that our list of binaries suffers from two incompleteness biases that should be lifted by Gaia DR4. Given the time spanned by the Gaia DR3 observations, no orbital solutions with periods in excess of 1400~d are present in the Gaia DR3 orbital catalogs, and as a consequence, these longer-period binaries appear as ‘acceleration solutions’ or ‘RV-trend’ in Gaia DR3. In the absence of any orbital elements for these, they cannot be screened as we did for the orbital solutions. For the Circular Cavity, they amount to 5191 astrometric ‘acceleration solutions’ and 532 ‘RV trends’, summing up to 5723. These numbers must be compared to 788 AstroSpectroSB1, 2013 astrometric, and 2109 SB1 orbits, summing up to 4910. For the Oval Cavity, there are to 2447 astrometric ‘acceleration solutions’ and 286 ‘RV trends’, summing up to 2773, as compared to 388 AstroSpectroSB1, 1185 astrometric, and 900 SB1 orbits, summing up to 2473.


\section{Discussion}

Figures 1-3 show that the gas of the approaching (negative velocity) side of the Circular Cavity - the LLIV Arch - is clumpy and knotty. The pre-supernova mass loss from the invisible companion of a spectroscopic binary was almost certainly clumpy \citep{Hamann2008, Smith2014}. Some of the features of the expanding shell may very well be a collection of these surviving clumps that rode the blast wave to achieve their observed high velocities. Recent evidence of this clumpy and knotty mass loss is seen in the spectacular new image of Wolf-Rayet 124 from JWST. Wolf-Rayet stars are prolific mass ejectors, shedding their outer layers and producing a halo of gas and dust. WR 124 itself has shed more than 10 M$_\odot$ to date, and the new image reveals that this ejected material is dominated by clumps and knots of gas and dust. 

One of these clump ejections was observed in real-time during the Great Dimming of Betelgeuse \citep{Dupree2022}. A substantial surface mass ejection started with a photospheric shock \citep{Kravchenko2021} and resulted in dust production in the atmosphere. Observations across the spectrum indicate that the ejected mass could represent a significant fraction of the annual mass-loss rate. Betelgeuse undergoes substantial mass loss via a stellar wind, and clumps and knots of dust in its environment suggest that material has been ejected in past events \citep{Kervella2011, Humphreys2022}.These phenomena indicated that mass loss in Betelgeuse may result from both a continuous wind and eruptive events.

\citet{Humphreys2022} find that optical and infrared imaging, spectra, and light curves provide clear observational evidence for discrete, directed gaseous outflows in the red hypergiant, VY CMa, as well as more typical red supergiants like Betelgeuse. Infrared-bright clumps and knots indicate that episodic events dominate the overall mass-loss rate from VY CMa and contribute significantly to the rate for typical red supergiants. Like the Sun, which ejects mass both continuously in the form of the solar wind and episodically via coronal mass ejections, supergiants also expel mass both continuously and episodically, but the winds are much more powerful and the ejections are much more massive. \citet{Humphreys2022} conclude that episodic outflows, related to magnetic fields and surface activity, are a major contributor to mass loss from red supergiants.

Figures 1-3 show that the appearance and behavior of the gas on the receding (positive velocity) side of the Circular Cavity are very different. This is the location of the NCP Loop \citep{Heiles1989, Meyerdierks1991}, two high-latitude cirrus complexes \citep{Barriault2010}, and several high-latitude molecular clouds \citep{Magnani1985}. The Oval Cavity seen in the last few panels of Fig.1 and centered at ($l$, $b$) = (134$^{\circ}$, 31$^{\circ}$) is traditionally associated with these well-studied features.

\citet{Meyerdierks1991} used FIR, HI, radio continuum, and soft X-ray data to test various expansion models and found that cylindrical expansion best matched the HI radial velocities. This result argues in favor of a distributed (rather than a point) source of expansion energy. Building on this work, \citet{Pound1997} used CO data to study the kinematics of the Ursa Major molecular clouds. Their results indicate that the clouds are located on the far side of an expanding bubble associated with the NCP Loop. The model is able to explain the gas velocity and line-width gradients, the centroid velocity shift between atomic and molecular features, and the large-scale IRAS color variations. \citet{Marchal2023} used HI data and dust extinction results to describe the elongated Oval Cavity that forms the inner part of the NCP Loop may be a protrusion of the Local Bubble. They model of the Oval Cavity as a prolate spheroid oriented toward the observer. They find that the components of the NCP Loop are 310 to 450 pc away.

The morphology of the NCP Loop resembles a blister on the edge of an HII region or a prominence in the solar atmosphere. It is tempting to support this analogy with Zeeman measurements of enhanced magnetic field strength in complexes along the NCP Loop \citep[see, e.g.,][]{Heiles1989, Myers1995}, but unfortunately, these results have been found to be spurious; the meticulous analysis of \citet{Verschuur1995} found that contamination from sidelobes mimics the Zeeman signature in HI emission profiles resulting in upper limits of the magnetic field that are much lower than the previously published values. 

\citet{Meyerdierks1991} used data from a series of seven rocket flights \citep{McCammon1983} to measure enhanced soft X-ray count rates in the central region of the NCP Loop, implying that the cavity is filled with hot plasma. Although these properties strongly imply a probable supernova origin, subsequent authors have shied away from this explanation noting that, (1) no OB stars are observed within the NCP Loop \citep{Humphreys1978}, and (2) the high galactic latitude where OB associations are not common. The non-spherical geometry of the cavity also argued against a single point-like energy source.

Figures 1-3 show that the appearance and behavior of the left and right sides of the receding (positive velocity) gas of the Circular Cavity are very different. Figure 2a shows that the gas on the left side of the Circular Cavity (higher galactic longitudes) recedes as expected from an expanding shell, forming a ridge at about $+$13 km s$^{-1}$ between $b$ = 20$^{\circ}$ and 38$^{\circ}$. The expansion of the gas on the right side of the Circular Cavity (lower galactic longitudes), however, is disrupted and not expanding freely. The morphology revealed by the $HI4PI$ data cube ($l$, $b$, $V_R$) indicates that this gas appears to be interacting with the HI approaching us on the near side of the Oval Cavity. The NCP Loop is at this intersection. This result (as well as information from the literature on the Oval Cavity) allows us to choose a preferred candidate from the binary systems listed in Tables 2 -- 4, namely Gaia DR3 1093757200530267520 (System \#25), as the center of the Circular Cavity. 

System \#25 belongs to category (i) in the list in \S2.4. The companion, with mass $1.34 \pm 0.32$~M$_\odot$, appears to be much more massive than the 0.7~M$_\odot$ primary. Could the 1.3~M$_\odot$ companion be a pair of twin 0.65~M$_\odot$ main-sequence stars instead of a single 1.3~M$_\odot$ neutron star? Assuming for the sake of simplicity that each of the two 0.65~M$_\odot$ companion stars has the same luminosity as the 0.7~M$_\odot$ primary, the twin secondary pair would be 0.75~mag \textit{brighter} than the primary, which is incompatible with both the SB1 (rather than SB2) nature of the system and the primary - secondary hierarchy. The system is located close to the approximate center of the Circular Cavity with a distance of 207~pc. 

Fig. 4 shows the $HI4PI$ map similar to Fig.~1k with the Circular Cavity in red and the Oval Cavity in blue. Neutron-star candidates are plotted as red diamonds with numbers corresponding to entries in Tables 2 -- 4. The result described above - that the Circular Cavity is in front of the Oval Cavity - as well as information from the literature on the distance and size of the Oval Cavity allows us to choose preferred candidates from the binary systems listed in Tables 2 -- 4. We’ve already seen that System \#25 at a distance of 207 pc may be responsible for the dynamics of the Circular Cavity, including the LLIV Arch. Gaia DR3 1144019690966028928 (System \#13) may be (partially) responsible for the Oval Cavity.

System \#13 hosts an excellent neutron-star candidate near the center of the Oval Cavity. The most convincing evidence comes from the masses ($M_1 = 0.74\pm0.05$~M$_\odot$, $M_2 = 1.39\pm0.12$~M$_\odot$) and the absence of light from the companion, which makes it an AMRF Class III system in the NSS classification by \citet{GaiaDR3-NSS}. All relevant quantities ($M_1, M_2, V_{\mathrm {cm}}, V_{\mathrm {space}}$; see Table~3) of system \#13 are surprisingly similar to those of System \#25. System \#13 is located near the center of the Oval Cavity at a distance of 488 pc. Moreover, with an orbital period of $1401 \pm 62$~d and an eccentricity of $0.38\pm0.02$, the orbital properties of System \#13 are similar to those of Gaia BH2 \citep{ElBadry2023-RGBH} .

There are two additional serious neutron-star candidates in the field of the Oval Cavity. The first is Gaia DR3 1131316620813278464 (System \#11), even though the companion mass seems a bit small ($1.17 \pm 0.05$~M$_\odot$). Although neutron stars have been observed with masses as low as 0.77~M$_\odot$ \citep{Doroshenko2022}, a massive white dwarf is more likely given the initial-mass function. The second is Gaia DR3 1114090365983444480 (System \#18), which is very likely to host a neutron star since the inferred mass is $1.25 \pm 0.12$~M$_\odot$. Multiple centers of expansion may account for the extended shape of the Oval Cavity along the line of sight.

Fig. 5 shows a polar projection of this scenario at $l$ = 135$^{\circ}$ including the intersection of the Circular Cavity (solid) and the Oval Cavity (dashed). As in Fig. 4, neutron-star candidates are plotted as red diamonds with numbers corresponding to entries in Tables 2 -- 4. The size, shape, and distance of the Oval Cavity are taken from \citet{Marchal2023} who model it as a prolate spheroid with a distance of 400 pc and a semi-minor and semi-major axes of 55 pc and 250 pc, respectively. The faint continuation of the ellipse inside the circle shows the extent of the Oval Cavity if it were not interacting with the Circular Cavity; this cut-off may help us understand why \citet{Meyerdierks1991} found that cylindrical expansion best matched the HI radial velocities. The oval in Fig.~4 may be the cross-section of this spheroidal or cylindrical feature in the plane of the sky. The red and blue components of the circle in Fig.~5 depict the locations of the NCP Loop and the lower arc, respectively, where the Circular Cavity and the Oval Cavity are interacting.

If the faint companions in the systems listed in Tables 2 -- 4 are indeed neutron stars, their progenitors must have been massive stars with $M_{\rm init} > 8$~M$_{\odot}$. Such massive stars are not expected to be found at large $|z|$ values, far from the Galactic plane. As a confirmation of this criterion, High-Mass X-Ray Binaries containing black holes are restricted to Galactic heights of up to 400 pc \citep{Fortin2022}. Therefore, a further criterion based on $|z| \le 400$~pc would exclude all candidates located farther away than about 700~pc (considering $b = 35^\circ$ as typical for our sample), which interestingly, corresponds to the farthest boundary of the Oval Cavity as mentioned above (Fig.~5). This criterion thus removes systems \#1, 2, 4, 6, 7, 10, 14, 15, 19, 21, 26, 30, and 33 from our sample of candidate neutron star. Italic typeface identifies them in Tables 2 -- 4. 

Using System \#25 to set the scale gives us a distance to the NCP loop of about 220~pc. \citet{Marchal2023}, however, find distances of about 300-350 pc along various lines of sight to features associated with the NCP Loop. Given that distance is one of the most challenging fundamental parameters to measure in astronomy, it is important to consider the uncertainties associated with each result.

The uncertainties on the distance to System \#25 are small, less than half a parsec, but the identification of the unseen companion as the center of the explosion that accelerated the LLIV Arch gas requires confirmation. We describe the unseen companion as a neutron-star $candidate$. 
Nevertheless, the $HI4PI$ data show definitively that an explosion did occur. System \#25 is either the cause of that explosion or simply the catalyst that motivated us to examine the surrounding HI gas in great detail.

The uncertainties on the distance of the NCP Loop from extinction, however, could be much larger. This distance and uncertainty analysis is described by \citet{Leike2020}, who used Gaia DR2 parallax and G-band photometric data combined with 2MASS, Pan-STARRS, and ALLWISE photometry to create a 3D dust extinction density data cube centered on the Sun with a distance spacing of 1 pc. Their maps show tendrils and filaments of dust on scales as small as their effective resolution of 2 pc. Their Figures 5 and 6 show the distances and uncertainties to the nearest and densest dust clouds, respectively. The results toward the NCP Loop show distances of about 300-350 pc, consistent with the results of \citet{Marchal2023}, but also uncertainties on these distances of tens of parsecs or greater, consistent with the distances anchored by System \#25 shown in Fig. 5. 

If the supernova of the invisible companion of a spectroscopic binary accelerated the anomalous-velocity gas and led to the other observational properties described in \S 2, then the distance to the star gives us a means to calculate some of the relevant physical parameters. The radius of the Circular Cavity in Fig. 1 is about 11$^{\circ}$ or 40 pc in latitude and somewhat larger in longitude. The gas of the approaching shell (the LLIV Arch) is about 167 pc away. This supernova is in the momentum-conserving snowplow phase where the dense shell continues to expand due to its own momentum and the interior continues to cool. A lower limit on the age of the Circular Cavity is about 100,000 years, enough time for the optical nebulosity and synchrotron radiation of the associated supernova to fade. We can estimate an upper limit if we make some simplifying assumptions - if the expansion velocity is constant and spherically symmetric, then the supernova took place about 700,000 years ago.

Figures 1-3 show that the expansion of the Circular Cavity is different on the approaching (negative velocity) side and the receding (positive velocity) side. The approaching gas is expanding into the low-density Local Bubble/Local Chimney \citep{Zucker2022, Lallement2003} whereas the receding gas is interacting with the denser low-velocity interstellar medium. Therefore, in order to estimate the energy of the original explosion, we use the near-side hemisphere and double it to account for the full, non-symmetric expansion. 

The $HI4PI$ data allow us to estimate the average column density of HI in a segment of the near side of the shell defined by the emission peaks at about -55 km s$^{-1}$ seen in Figs. 2 \& 3. A composite spectrum for the area bounded by longitudes 120$^{\circ}$ \& 172$^{\circ}$ and latitudes 24$^{\circ}$ \& 45$^{\circ}$ produced a single profile, and a Gaussian fit to the negative velocity component indicates an HI column density of 2.5 $\times$ 10$^{19}$ cm$^{-2}$. At the distance of 167 pc, this implies a mass of about 1000 M$_\odot$ for the LLIV Arch. Doubling this to account for the receding side of the Circular Cavity gives us 2000 M$_\odot$. Multiplying this by the observed LLIV Arch velocity results in a momentum, p $=$110,000 M$_\odot$ km s$^{-1}$, in good agreement with recent model estimates for a supernova with an age greater than 100,000 yrs \citep[e.g.,][see their Figs. 3, 5-6 and 8]{Haid2016}.

If one or more of the binaries in Fig. 5 host a neutron star (the most promising candidates are Systems \#11, 13, and 25) that underwent a supernova explosion, then they can account for the extended shape of the Oval Cavity along the line of sight and anchor the distances to the LLIV Arch and NCP Loop. If they are not neutron stars, then the distances and the associated uncertainties may a bit larger, and the scenario shown in Fig. 5 simply scales to the new distance. 

\section{Conclusions}

Supernova explosions attributed to the unseen companion in several binary systems identified by Gaia DR3 may be responsible for a number of well-known and well-studied features in the radio sky including the LLIV Arch and the NCP Loop. 

Slices from the Longitude-Latitude-Velocity data cube of the $\lambda$-21-cm galactic neutral atomic hydrogen $HI4PI$ survey show multiple signatures of an expanding shell. 

The Circular Cavity is best seen at low velocities, centered near System \#25 with a radius of about 11$^{\circ}$ (40 pc) and a distance of 207 pc. 

The NCP Loop is located on the rim of the Circular Cavity, and the gas of the approaching shell is part of the LLIV Arch. 

If we make the simplifying assumptions that the expansion of the Circular Cavity is uniform and spherically symmetric, then the explosion took place about 700,000 years ago. The momentum, p $=$110,000 M$_\odot$ km s$^{-1}$, is in reasonable agreement with recent model estimates for a supernova with this age.

The morphology revealed by the $HI4PI$ data cube indicates that the gas on the receding (positive velocity) side of the Circular Cavity appears to be interacting with the HI approaching us on the near side of the Oval Cavity; the NCP Loop is at this intersection (Fig. 4).

The neutron star candidate in the System \#13 binary may be (in part) responsible for this Oval Cavity. 

If two or more binaries in the Oval Cavity host a neutron star that underwent a supernova explosion, then they can account for the extended shape along the line of sight; the most promising candidates are Systems \#11, 13, and 18.

We can use the primary star in these binaries to anchor the distances to the LLIV Arch and the NCP Loop, which are about 167 and 220 pc, respectively.

The distance to the NCP Loop from extinction is larger, but the uncertainty is also larger, tens of parsecs or greater.

Given that distance is one of the most challenging fundamental parameters to measure in astronomy, a detailed analysis of the extinction uncertainties would be a worthy contribution to the extensive body of literature on the well-known and well-studied features - LLIV Arch, NCP Loop, high-latitude cirrus complexes, high-latitude molecular clouds – that are associated with the Circular Cavity and the Oval Cavity. 

\section{Acknowledgments}

We are grateful to T. Dame for providing his MacFits software and J. Kerp for sending the $HI4PI$ data. We would also like to thank W.B. Burton, J. Raymond, and S. Reynolds for helpful advice. A.J. is partly supported by FNRS-F.R.S. research project PDR T.0115.23.

This work presents results from the European Space Agency (ESA) space mission Gaia. Gaia data are processed by
the Gaia Data Processing and Analysis Consortium (DPAC). Funding for the
DPAC is provided by national institutions, in particular the institutions participating in the Gaia MultiLateral Agreement (MLA). The Gaia mission website is \url{https://www.cosmos.esa.int/gaia}. The Gaia archive website is \url{https://archives.esac.esa.int/gaia}.

\section{References}
\bibliographystyle{aa} 
\bibliography{references} 

\clearpage

\renewcommand{\arraystretch}{1.5}
\renewcommand{\tabcolsep}{4pt}

\begin{table}
\caption{
\label{Tab:fM}
Maximum value of the spectroscopic mass function $f(M_1,M_2)$ as a function of the component masses $M_1, M_2$ (in M$_\odot$). Entries in boldface correspond to max $f(M_1,M_2) \ge 0.3$~M$_\odot$}
\vspace{2mm}
\begin{center}
{\scriptsize
\begin{tabular}{r|cccccccc}
\hline
\phantom{$M_1$} & $M_2$   &1.1&1.2&1.3&1.4&1.5&1.6\\
$M_1$   &  \\
\hline
0.5     &         &0.213&0.276&{\bf 0.352}&{\bf 0.439}&{\bf 0.540}&{\bf 0.655}\\
0.6     &         &0.197&0.256&{\bf 0.325}&{\bf 0.406}&{\bf 0.499}&{\bf 0.606}\\
0.7     &         &0.183&0.237&{\bf 0.301}&{\bf 0.376}&{\bf 0.463}&{\bf 0.562}\\
0.8     &         &0.170&0.220&0.280&{\bf 0.350}&{\bf 0.430}&{\bf 0.522}\\
0.9     &         &0.158&0.205&0.261&{\bf 0.326}&{\bf 0.401}&{\bf 0.487}\\
1.0     &         &0.148&0.192&0.244&{\bf 0.305}&{\bf 0.375}&{\bf 0.455}\\
1.1     &         &0.139&0.180&0.229&0.286&{\bf 0.351}&{\bf 0.426}\\
1.2     &         &0.130&0.169&0.215&0.268&{\bf 0.330}&{\bf 0.400}\\
1.3     &         &0.122&0.159&0.202&0.252&{\bf 0.310}&{\bf 0.376}\\
\hline
\end{tabular}
}
\end{center}
\end{table}

\begin{table}
\caption{
\label{Tab:coords}
Promising candidate systems that could host a neutron-star companion{$^{a,b}$} }
\begin{center}
{\scriptsize
\begin{tabular}{>{\rowmac}c>{\rowmac}c>{\rowmac}c>{\rowmac}c>{\rowmac}c>{\rowmac}c>{\rowmac}c>{\rowmac}c>{\rowmac}c>{\rowmac}c<{\clearrow}}
\hline
\# & Gaia DR3 ID & NSS type & $\alpha$ [h m s] & $\delta$ [$^{\circ}$ ' ''] & $l$ [$^{\circ}$] & $b$ [$^{\circ}$] & $\varpi$ [mas] & Dist. [pc] & $|z|$ [pc] \\
\hline
\setrow{\itshape}
1& 
1717768497839377920 & Astrometric  & 12 39 24.86 & +79 56 57.89 & 123.59 & 37.16 &  $0.80\pm 0.02$  & 1250&755 \\
\setrow{\itshape}
2 & 1720105647243105024 & Astrometric  & 12 32 42.13 & +82 21 11.20 & 123.69 & 34.75 &  $0.98 \pm 0.02$ & 1020&581 \\
\setrow{\upshape}
3 & 1719340593308938112 & Astrometric & 12 32 44.39 & +80 28 59.03  & 123.89 & 36.61 & $1.85\pm0.01$    & 541 &323 \\
\setrow{\itshape}
4 & 1133382981119378688 & Astrometric & 11 33 17.37 & +80 53 54.72 & 126.66 & 35.66 &  $1.14\pm0.04$   & 877 &511 \\
\setrow{\upshape}
5 & 1134341411661382912 & Astrometric  & 10 43 17.85 & +82 30 59.43 & 127.68 & 33.39 &  $1.50\pm0.02$   & 667 &367  \\
\setrow{\itshape}
6 & 1129046129301663744 & SB1          & 11 32 55.22 & +78 19 13.34 & 127.89 & 38.04 & $1.12\pm 0.03$   & 893 &550\\
\setrow{\itshape}
7 & 1129693294973971456 & Astrometric  & 10 46 58.31&+78 46 15.39   & 130.13 & 36.56 & $0.76\pm0.03$ &    1316&784\\
\setrow{\upshape}
8 & 1145598142986710400 & Astrometric & 08 25 30.48 & +82 31 52.92  & 130.83 & 29.89 & $2.22\pm0.01$ &    450 &224\\
9 & 1148571939686970752 & Astrometric & 08 11 16.16 & +82 15 29.35  & 131.30 & 29.52 &  $1.71\pm0.04$ &   585 &288\\
\setrow{\itshape}
10 & 1131412278324875264& AstroSpectroSB1 & 10 04 41.03 &+78 40 57.64&132.13 & 35.28 & $1.28\pm 0.02$ &   781 &451\\
\setrow{\upshape}
11 & 1131316620813278464 & AstroSpectroSB1 &10 03 13.62 & +78 32 52.27&132.31& 35.31 & $4.09\pm0.01$ &    244 &141\\
12 & 1130964845812456064 & Astrometric & 09 45 29.66 & +78 10 13.16 &  133.37& 34.88 &  $4.33\pm0.02$ &    231&132 \\
13 & 1144019690966028928 & Astrometric & 08 53 25.15 &	+79 21 18.48 & 133.76& 32.11 & $2.05\pm0.01$ &    488 &259\\
\setrow{\itshape}
14 & 1124657153762302336& Astrometric &09 15 17.61 &+75 45 39.48 & 136.96 &	   34.79 & $1.08\pm 0.01$ &   926 &528\\
\setrow{\itshape}
15 & 1058875159778407808 & Astrometric & 10 48 59.45 & +65 47 55.84 & 140.71 & 46.88 & $0.90\pm0.03$ &    1111&811\\
\setrow{\upshape}
16 & 1120612943836509568 & Astrometric & 09 09 27.71 & +72 30 11.25 & 140.74 &  35.91 & $1.73\pm0.01$ &    578 &339\\
17 & 1112080286929477376 & Astrometric & 07 28 16.25 & +74 07 18.42 & 140.83 & 28.53 &  $1.31\pm 0.07$ &  763 &364\\
18 & 1114090365983444480 & Astrometric & 06 28 59.66 &	+72 45 48.08 & 141.79& 24.19 & $1.10\pm0.03$ &    909 &372 \\ 
\setrow{\itshape}
19 & 1111932402615746432 & Astrometric & 07 26 22.69 & +73 12 33.93 & 141.87 & 28.40 & $1.01\pm0.02$ &    990 &471 \\
\setrow{\upshape}
20 & 1107717287351599488 & Astrometric & 06 04 19.88 & +71 08 41.75 & 142.87 & 21.89 &  $2.01\pm 0.05$ &  498 &186\\
\setrow{\itshape}
21 & 1111328358414543616	&	SB1 & 07 22 03.00 &	+71 49 12.70 & 143.44 &    28.05 & $0.98\pm0.02$ &    1020&480 \\
\setrow{\upshape}
22 & 1097717538333601408 & Astrometric+SB1  & 08 19 53.23 & 	+70 12 35.49 & 144.86& 32.85 & $10.88\pm 0.02$ &  91.9&50 \\
23 & 1110185721018656384 & SB1 & 07 44 30.15 & +70 05 39.45 & 145.40 &         29.89 & $1.99\pm 0.02$ &   503 &251  \\
24 & 1094277819285305856 & AstroSpectroSB1 &08 38 18.69 & +68 38 38.06&146.26& 34.78 &$2.99\pm 0.03$ &    334 &191 \\
25 & 1093757200530267520 & AstroSpectroSB1 &08 23 34.58&+67 14 52.16&148.27 &  33.72 &$4.83\pm0.01$&      207 &115  \\
\setrow{\itshape}
26 & 1093755272088380032	& SB1      & 08 23 30.69	&+67 10 42.50 &148.35& 33.73 & $0.97\pm0.04$ &    1031&572 \\
\setrow{\upshape}
27 & 1095948870802554368 & Astrometric & 07 38 14.68 & +66 57 22.04 & 149.02 & 29.36 & $1.36\pm0.02$ &    735 &360  \\
28 & 1100973226624010240 & Astrometric &06 44 13.72 &	+65 14 02.41 & 150.15& 23.72 & $2.59\pm0.03$ &    386 &155\\
29 & 1043525457762566784 & AstroSpectroSB1 &09 02 51.68 &+63 33 07.80 &151.52& 38.56 & $1.64\pm0.02$ &    610 &380 \\
\setrow{\itshape}
30 & 1089293007226401152 & SB1 & 07 25 42.43 &	+64 32 31.59 &         151.69& 27.95 & $1.07\pm 0.02$ &   935 &438  \\
\setrow{\upshape}
31 & 1088587331214903552 & Astrometric+SB1  & 07 43 04.55 & +63 42 25.67 & 152.77 &         29.79 & $1.52\pm 0.07$ &   658 &327 \\
32 & 1086410878011578240 & Astrometric+SB1 &07 37 40.60 &+61 12 37.43&155.60 & 29.05 & $2.94\pm0.04$ &    340 &165  \\
\setrow{\itshape}
33 & 1002740757557931264 & Astrometric & 07 03 33.33 &+60 04 41.25 & 156.20 &  24.75 & $0.58\pm0.02$ &    1724&722 \\ 
\setrow{\upshape}
34 & 1036462439880032000 & Astrometric+SB1&09 05 40.26 &+56 19 59.78 &160.48 & 40.71 &$2.72\pm 0.02$ &    367 &239 \\
\\
\hline
\multicolumn{9}{c}{Gaia BH}\\
\hline
&Gaia BH1 & Astrometric & 17 28 41.09 & -00 34 51.52 & 22.63 & 18.05 & $2.10\pm0.02$& 476 & 148\\
&Gaia BH2 & AstroSpectroSB1 & 13 50 16.75 & -59 14 20.33 & 310.40 & 2.78 & $0.86\pm0.02$& 1163 & 79\\
\hline\\
\end{tabular}
}
\end{center}
{\scriptsize
{$^a$}Sorted by increasing Galactic longitude.\\
{$^b$}Rows in italic font identify systems located more than 400~pc above the Galactic plane.\\
}

\end{table}

\clearpage
\clearrow

\renewcommand{\tabcolsep}{3pt}
\begin{table}
\caption{
\label{Tab:NSSinfo}
Promising candidate systems and values important for our selection criteria. \\
}
\begin{center}
{\scriptsize
\begin{tabular}{>{\rowmac}c>{\rowmac}c>{\rowmac}c>{\rowmac}c>{\rowmac}c>{\rowmac}c>{\rowmac}c>{\rowmac}c>{\rowmac}c>{\rowmac}c>{\rowmac}c>{\rowmac}c>{\rowmac}c>{\rowmac}c>{\rowmac}c>{\rowmac}c<{\clearrow}}
\hline
\# & Gaia DR3 ID & NSS type & $M_1$ & $M_{2[,\mathrm {lower}]}^a$ & $f(M)$ & AMRF & \multicolumn{1}{c}{$V_{\mathrm{cm}}$}  &\multicolumn{1}{c}{$V_{\mathrm{space}}$}  & \multicolumn{1}{c}{$P$} & \multicolumn{1}{c}{$e$} &\multicolumn{4}{c}{Distance from center} & \multicolumn{1}{c}{Rem.}\\
\cline{12-15}&&&&(M$_\odot$)&(M$_\odot$)&(M$_\odot$)&\multicolumn{1}{c}{(km/s)}&\multicolumn{1}{c}{(km/s)}& (d)& &\multicolumn{2}{c}{Circular}&\multicolumn{2}{c}{Oval}\\
&&&&&&&&&&&now & $t_{\rm SN}$ & now & $t_{\rm SN}$ \\
&&&&&&&&&&&($^\circ$)&($^\circ$)&($^\circ$)&($^\circ$) \\
\hline
\setrow{\itshape}
1 & 1717768497839377920 & Astrometric & 1.0 & $1.15\pm0.20$ &  - & 0.70       & -37.6& 38.8 & 1029 & 0.35&17.7&17.7&9.1 &9.3\\
2 & 1720105647243105024 & Astrometric  & 1.0 & 0.76 - 1.71 & - & 0.61         & -    &  -   & 763  & 0.82&17.6&17.6&8.1 &8.2\\
\setrow{\upshape}
3 & 1719340593308938112 & Astrometric & 1.0 & 1.10 - 1.35 & - & 0.37          & -2.6& 39.7 & 233  & 0.38&17.4&17.4&8.6 &8.6\\
\setrow{\itshape}
4 & 1133382981119378688 & Astrometric & 0.8 & 0.81 - 1.79 & - & 0.54          & -   & -    & 712  & 0.57&15.1&15.2&6.3 &6.3\\
\setrow{\upshape}
5 & 1134341411661382912 & Astrometric  &1.1 & 0.47 - 1.38 & - & 0.37          & -22.6& 51.8 & 321  & 0.54&14.4&14.4&4.7 &4.7\\
\setrow{\itshape}
6 & 1129046129301663744 & SB1 & 1.5 & 1.34 & 0.40 & -                        & 14.7& -    & 1.6  & 0.0 &14.4&14.5&7.3 &7.3\\
7 & 1129693294973971456 & Astrometric & 1.3 & $1.27\pm0.13$ & - & 0.63        & -38.3& 73.4 & 956  & 0.34&12.4&12.3&5.1 &5.2\\
\setrow{\upshape}
8 & 1145598142986710400 & Astrometric & 0.9 & 0.56 - 3.15  & - & 0.58         & -32.5& 44.5 & 2803 & 0.63&12.7&12.6&2.8 &2.7\\
9 & 1148571939686970752 & Astrometric & 1.5 & 0.40 - 1.65 & - & 0.24          & 3.5& 32.9 & 622  & 0.18&12.5&12.6&2.9 &3.0\\ 
\setrow{\itshape}
10 & 1131412278324875264 & AstroSpectroSB1 & - & - & 0.49 & -                 & 17.6& 33.4 & 923  &0.008&10.7&10.6 &3.4 &3.2\\
\setrow{\upshape}
11 & 1131316620813278464 & AstroSpectroSB1 & 1.1 &$1.17\pm0.05$ & 0.30 & 0.62 & -32.2& 38.3 & 479  & 0.27&10.5&10.6&3.4 &3.2\\
12 & 1130964845812456064 & Astrometric & 0.7 & 0.52 - 1.68 & - & 0.63         & 24.4& 59.2 & 1343 & 0.18&9.6 &10.2&2.9 &2.2\\
13 & 1144019690966028928 & Astrometric & 0.7 & $1.39\pm0.12$ & - & 0.96       & -42.5& 80.2 & 1401 & 0.38&9.6 &8.8 &0.7 &1.4&AMRF ClassIII\\
\setrow{\itshape}
14 & 1124657153762302336 & Astrometric & 1.1 & $1.20\pm0.16$ & - & 0.68       & 44.7& 55.6 & 540  & 0.67&6.7 &6.8 &4.3 &4.2\\
15 & 1058875159778407808 & Astrometric & 1.09 & $1.78\pm0.15$ & - & 0.85      & 60.1& 88.2 & 835  & 0.42&13.3&13.1&15.8&15.7 & AMRF ClassIII\\
\setrow{\upshape}
16 & 1120612943836509568 & Astrometric & 1.0 & 1.1 - 1.8 & - & 0.62           & -26.6& 30.8 & 1229 & 0.45&4.0 &3.9 &7.4 &7.6\\
17 & 1112080286929477376 & Astrometric & 0.7 & 0.50 - 1.26 & - & 0.60         & -   & -    & 1046 & 0.23&6.5 &6.6 &7.7 &7.3\\
18 & 1114090365983444480 & Astrometric & 1.1 & $1.25\pm0.12$ & - & 0.68         & 25.5& 57.1 & 1058 & 0.47&10.2&10.2&11.2&11.7   & AMRF ClassIII\\
\setrow{\itshape}
19 & 1111932402615746432 & Astrometric &1.0&0.89 - 1.71 & - &0.67             & -24.7& 36.2  & 1223 & 0.41&6.2 &6.2 &8.6 &8.6\\
\setrow{\upshape}
20 & 1107717287351599488 & Astrometric & 0.6 & 0.66 - 1.60 & - & 0.51         & -    & -    & 1613 & 0.73&12.3&12.2&13.6&12.9 \\
\setrow{\itshape}
21 & 1111328358414543616 & SB1 & - & - & 0.31 & -                             & -41.0& -    & 141  & 0.16&6.1 &6.3 &10.0&9.4 \\
\setrow{\upshape}
22 & 1097717538333601408 & Astrometric+SB1$^b$ &  0.9& $1.23\pm0.08$&0.39 & 0.34 & -40.7& 42.0 & 45   & 0.37&1.5 &1.5 &10.0&9.4 \\
23 & 1110185721018656384 & SB1          &1.1&1.3& 0.43 & -                    & -1.1& -    & 49   & 0.28&4.1 &4.2 &11.0&10.8 \\
24 & 1094277819285305856 & AstroSpectroSB1 &1.3 & $1.49\pm0.38$ & 0.02 & 0.54 & 5.5& 27.0 & 1004 & 0.33&1.3 &0.9 &11.2&10.9 \\
25 & 1093757200530267520 & AstroSpectroSB1 &0.7&$1.34\pm0.32$&0.38& 0.37  & -40.7& 74.4 & 584  & 0.78&2.7 &2.9 &12.9&13.1 \\
\setrow{\itshape}
26 & 1093755272088380032 & SB1 & - & - & 0.42 & -                             & -21.4& -    & 335  & 0.13&2.8 &2.9 &12.9&13.1 \\
\setrow{\upshape}
27 & 1095948870802554368 & Astrometric &1.1&$1.15\pm0.05$& - & 0.64           & -29.6& 38.9 & 313  & 0.31&5.8 &5.9 &14.2&14.2 \\
28 & 1100973226624010240 & Astrometric & 0.7 & $1.28\pm0.14$ & - & 0.92       & -16.0& 23.8 & 1804 & 0.66&11.2&11.4&17.8&17.4  & AMRF ClassIII\\
29 & 1043525457762566784 & AstroSpectroSB1 &1.4&$1.49\pm0.18$&0.22& 0.58      & -44.5& 55.2 & 310  & 0.25&7.0 &7.0 &15.9&16.6 \\
\setrow{\itshape}
30 & 1089293007226401152 & SB1 & - & - & 0.34 & -                             & -38.5& -     & 45  & 0.17&8.3 &8.3 &17.0&16.6 \\
\setrow{\upshape}
31 & 1088587331214903552 & Astrometric+SB1$^b$ & - & - & 0.60 & -             & -46.3& 48.2  & 916 & 0.17&7.8 &7.8 &17.3&17.0 \\
32 & 1086410878011578240 & Astrometric+SB1$^b$ &1.1&$1.58\pm0.25$& 0.56 & 0.56& -7.3& 8.0   & 895 & 0.64&10.3&10.3&20.0&19.7  \\
\setrow{\itshape}
33 & 1002740757557931264 & Astrometric & 1.7 & $1.93\pm0.30$ & - & 0.70       & -23.0& 40.6  & 922 & 0.77&13.4&13.4&22.3&21.6   \\
\setrow{\upshape}
34 & 1036462439880032000 & Astrometric+SB1$^b$&0.9&$1.32\pm0.12$ & 0.40 & 0.35& -32.1& 36.5  & 229 & 0.32&14.0&14.2&22.6&23.9   \\ 
\hline
\multicolumn{13}{c}{Gaia BH}\\
\hline
&Gaia BH1 & Astrometric & 1.0 & 12.8 & 3.98 & 2.32 & 76 & 47 & 186 & 0.49 & - & - \\
&Gaia BH2 & AstroSpectroSB1 & 1.0 & 8.4 & 1.33 & 6.72 & -4 & 65 & 1276 & 0.52 & - & -\\
\hline\\
\end{tabular}
}
\end{center}
{\scriptsize
$^a$ Secondary mass and its error, the minimum secondary mass, or the possible mass range.\\
$^b$ Despite being seen as both astrometric and SB1 binaries, separate solutions were computed in Gaia DR3, contrarily to AstroSpectroSB1 cases.
}
\end{table}

\clearpage

\renewcommand{\arraystretch}{1.5}
\renewcommand{\tabcolsep}{5pt}
\begin{table}
\caption{
\label{Tab:stellarparams}
Stellar parameters of the primary stars. }
\begin{center}
{\scriptsize
\begin{tabular}{lccccccc}
\hline
\# & Gaia DR3 ID & $G$ [mag] & $M_G$ [mag] & $B_p - R_p$ [mag] & $T_{\rm eff}$ [K] & $\log g$ [dex] & $[M/H]$ [dex]
\medskip\\
\hline
1 & 1717768497839377920 &	14.71 &		3.98 &		0.91 &		5932&	3.8&	-0.52\\
2 & 1720105647243105024 &	14.96 &		4.91 &		0.95 &		5659&	4.5&	-0.09\\
3 & 1719340593308938112 &	12.95 &		4.26 &		0.97 &		5434&	4.2&	-0.33\\
4 & 1133382981119378688 &	15.57 &		6.00 &		1.30 &		\\
5 & 1134341411661382912 &	13.28 &		4.29 &		0.77 &		5951&	4.3&	-0.49\\
6 & 1129046129301663744 &	12.24 &		2.48 &		0.59 &		\\
7 & 1129693294973971456 &	13.84 &		2.68 &		0.66 &		6274&	3.9&	-0.75\\
8 & 1145598142986710400 &	13.63 &		5.67 &		1.00 &		\\
9 & 1148571939686970752 &	11.40 &		2.51 &		0.55 &		6163&	3.6&	-0.88\\
10 & 1131412278324875264 &	9.89 &		0.11 &		1.12 &		4791&	2.1&	-0.25\\
11 & 1131316620813278464 &	10.72 &		3.66 &		0.72 &		6434&	4.4&	-0.23\\
12 & 1130964845812456064 &	13.83 &		7.07 &		1.35 &		\\
13 & 1144019690966028928 &	13.57 &		4.38 &		0.71 &		6008&	4.5&	-0.41\\
14 & 1124657153762302336 &	13.91 &		3.96 &		0.78 &		5860&	4.2&	-0.84\\
15 & 1058875159778407808 &	14.52 &		5.02 &		0.75 &		5881&	4.6&	-0.19\\
16 & 1120612943836509568 &	13.77 &		5.42 &		1.07 &		\\
17 & 1112080286929477376 &	16.94 &		7.45 &		1.59 &		4600&	4.3&	0.50\\
18 & 1114090365983444480 &	14.25 &		4.14 &		0.85 &		6113&	3.7&	-0.87\\
19 & 1111932402615746432 &	14.58 &		5.02 &		0.85 &		5472&	4.3&	-0.98\\
20 & 1107717287351599488 &	16.46 &		7.73 &		2.14 &		6022&	4.5&	-1.17\\
21 & 1111328358414543616 &	10.85 &		0.80 &		1.03 &		\\
22 & 1097717538333601408 &	8.94 &		4.16 &		0.88 &		5650&	4.3&	-0.11\\
23 & 1110185721018656384 &	12.22 &		3.72 &		0.83 &		5941&	4.1&	-0.34\\
24 & 1094277819285305856 &	10.73 &		3.38 &		0.58 &		3175&	5.0&	0.00\\
25 & 1093757200530267520 &	12.84 &		6.19 &		1.35 &		4747&	3.9&	0.05\\
26 & 1093755272088380032 &	12.06 &		1.98 &		0.87 &		5827&	4.3&	-0.06\\
27 & 1095948870802554368 &	13.20 &		4.41 &		0.75 &		\\
28 & 1100973226624010240 &	14.70 &		7.78 &		1.18 &		4854&	4.6&	0.05\\
29 & 1043525457762566784 &	11.89 &		3.55 &		0.72 &		6592&	4.4&	-0.09\\
30 & 1089293007226401152 &	11.77 &		1.91 &		0.96 &		\\
31 & 1088587331214903552 &	9.45 &		0.36 &		1.07 &		5037&	2.5&	-0.10\\
32 & 1086410878011578240 &	11.32 &		3.34 &		0.80 &		6056&	4.2&	-0.25\\
33 & 1002740757557931264 &	13.23 &		1.99 &		0.64 &		6859&	3.8&	-0.18\\
34 & 1036462439880032000 &	12.15 &		4.33 &		0.90 &		5342&	5.0&	 \\
\\
\hline
\multicolumn{8}{c}{Gaia BH}\\
\hline
& Gaia BH1 & 13.77 & 4.52 & 1.18 & 5396.4$^a$ & 4.23$^a$ & -1.07$^a$ \\
& Gaia BH2 & 12.28 & 2.57 & 1.49 & 4831.7$^b$ & 3.25$^b$ & 0.01$^b$	\\
\hline\\
\end{tabular}
}
\end{center}
{\scriptsize
{$^a$}\citet{ElBadry2023-MSBH} find $T_{\rm eff} = 5850 \pm 50$~K, $\log g = 4.55 \pm 0.16$, and [Fe/H] ~$= -0.20 \pm 0.05$.\\
{$^b$}\citet{ElBadry2023-RGBH} find $T_{\rm eff} = 4604 \pm 87$~K, $\log g = 2.71 \pm 0.24$, and [Fe/H] ~$= -0.22 \pm 0.02$.\\
}
\end{table}

\clearpage

\begin{figure}
\figurenum{1}
\epsscale{1.}
\plotone{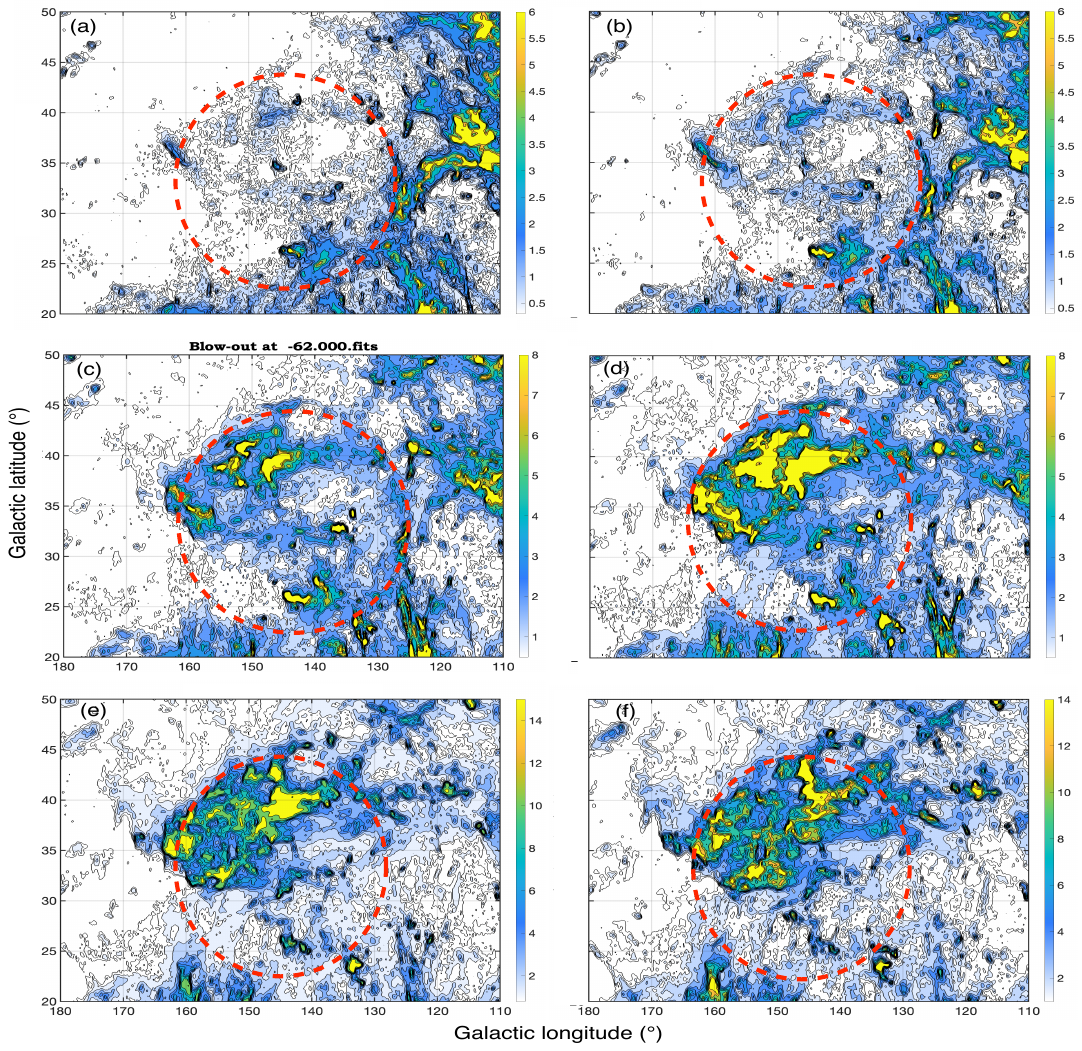}
\caption{Area maps centered showing the details of the environment at different velocities using $HI4PI$ data with a 2 km s$^{-1}$ band width. In each frame the red dashed circle highlights the approximate boundary of the low-velocity Circular Cavity. Panels (a) through (f) at -70, -66, -62, -58, -54, and -50 km s$^{-1}$, respectively, show different velocity slices of the approaching shell. Legends in Kelvins.
}
\end{figure}
\clearpage

\begin{figure}
\figurenum{1cont}
\epsscale{1.}
\plotone{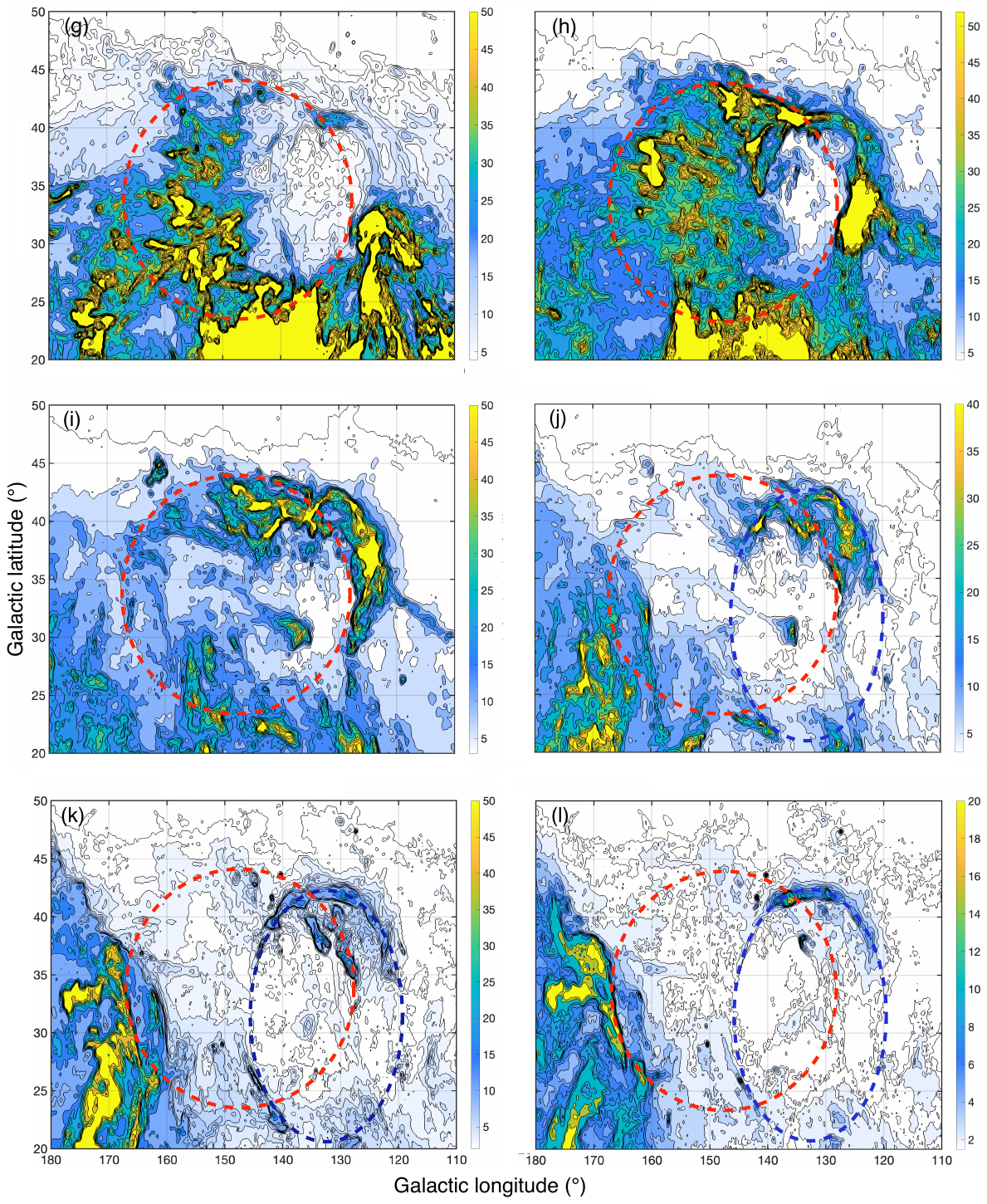}
\caption{(g) at -2 km s$^{-1}$ shows the NCP Loop as well as a circular structure with multiple extending threads in the lower-left quadrant. (h) at $+$2 km s$^{-1}$ shows the complex Circular Cavity structure and a more-defined image of the NCP Loop. (i) at $+$6 km s$^{-1}$ shows a circular structure with multiple extending threads in the upper-right quadrant, similar to the fainter structure seen in panel (g). (j) at $+$10 km s$^{-1}$ shows the first hints of the oval cavity (blue dash) traditionally associated with the NCP Loop. (k) $+$14 km s$^{-1}$ shows the lower arc at ($l$, $b$) = (143$^{\circ}$ , 25$^{\circ}$) that helps define the oval cavity. (l) at $+$18 km s$^{-1}$ higher-velocity traces of the Oval Cavity. Legends in Kelvins.
}
\end{figure}
\clearpage

\begin{figure}
\figurenum{2}
\epsscale{1.}
\plotone{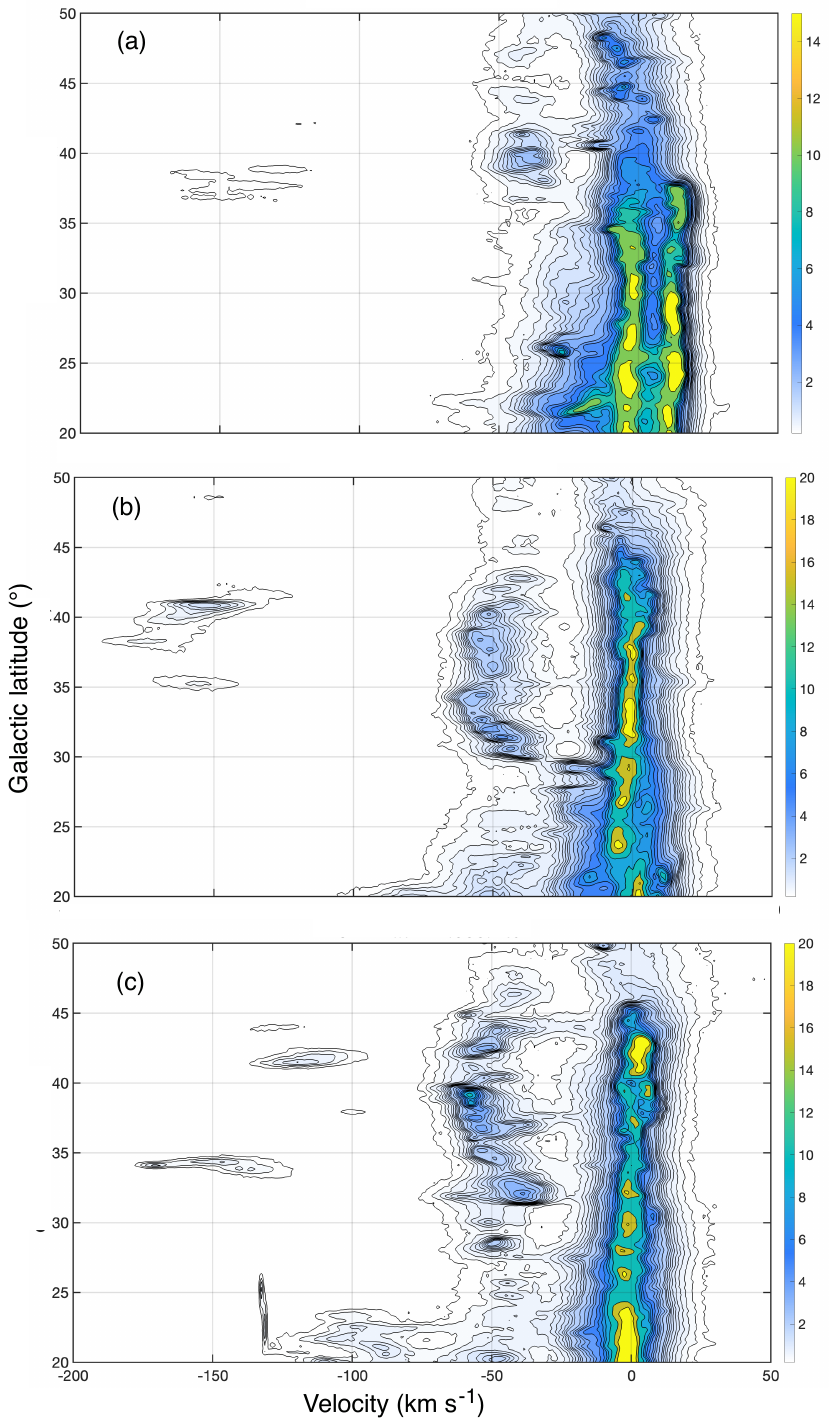}
\caption{($b$, $v$) maps at different longitudes. (a) at $l$ = 169$^{\circ}$ shows twin ridges of positive- and near zero-velocity gas. The positive-velocity ridge may represent the receding component of the expanding shell. (b) at $l$ = 157$^{\circ}$ shows the HI ridge at -55 km s$^{-1}$ known as the LLIV Arch with a velocity gradient consistent with an HI shell expanding away from the source and toward the observer. (c)  at $l$ = 147$^{\circ}$ shows that the shell structure is extremely clumpy. Legends in Kelvins.
}
\end{figure}
\clearpage

\begin{figure}
\figurenum{2cont}
\epsscale{1.}
\plotone{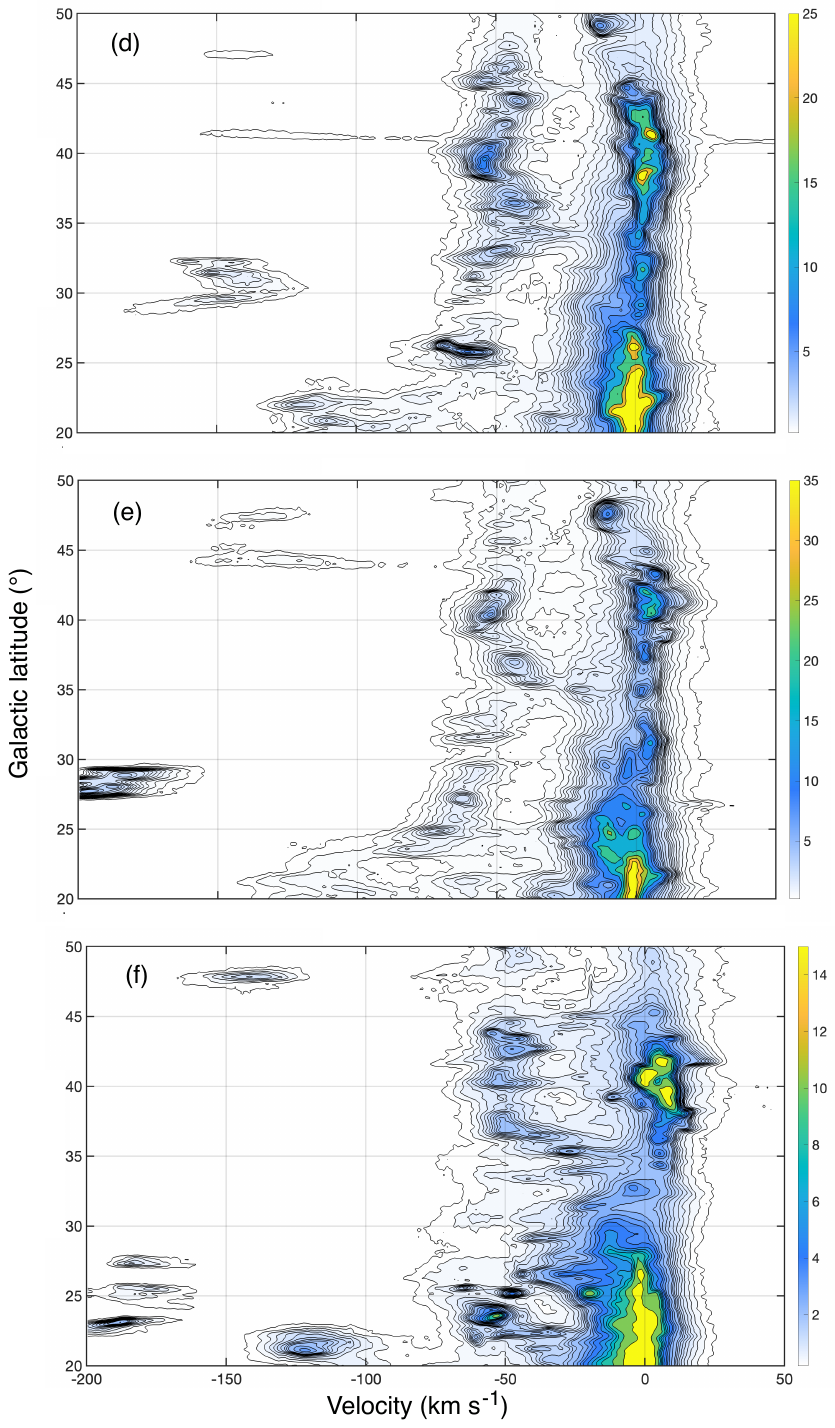}
\caption{(d) at 143$^{\circ}$ cuts across the center of the shell outlined in Fig. 1. The distortion of the low-velocity gas is consistent with expansion on the far side. (Note: The long, horizontal contour at 40.$^{\circ}$9 is the galaxy M 81.) (e) at 139$^{\circ}$, the near side of the shell shows knots and clumps. A connection to low-velocity gas is apparent at about 33$^{\circ}$ and 46$^{\circ}$. (f) at $l$ = 133$^{\circ}$, the bright structure at positive velocities up to $+$10 km s$^{-1}$ is the NCP Loop. Legends in Kelvins.
}
\end{figure}
\clearpage

\begin{figure}
\figurenum{3}
\epsscale{.9}
\plotone{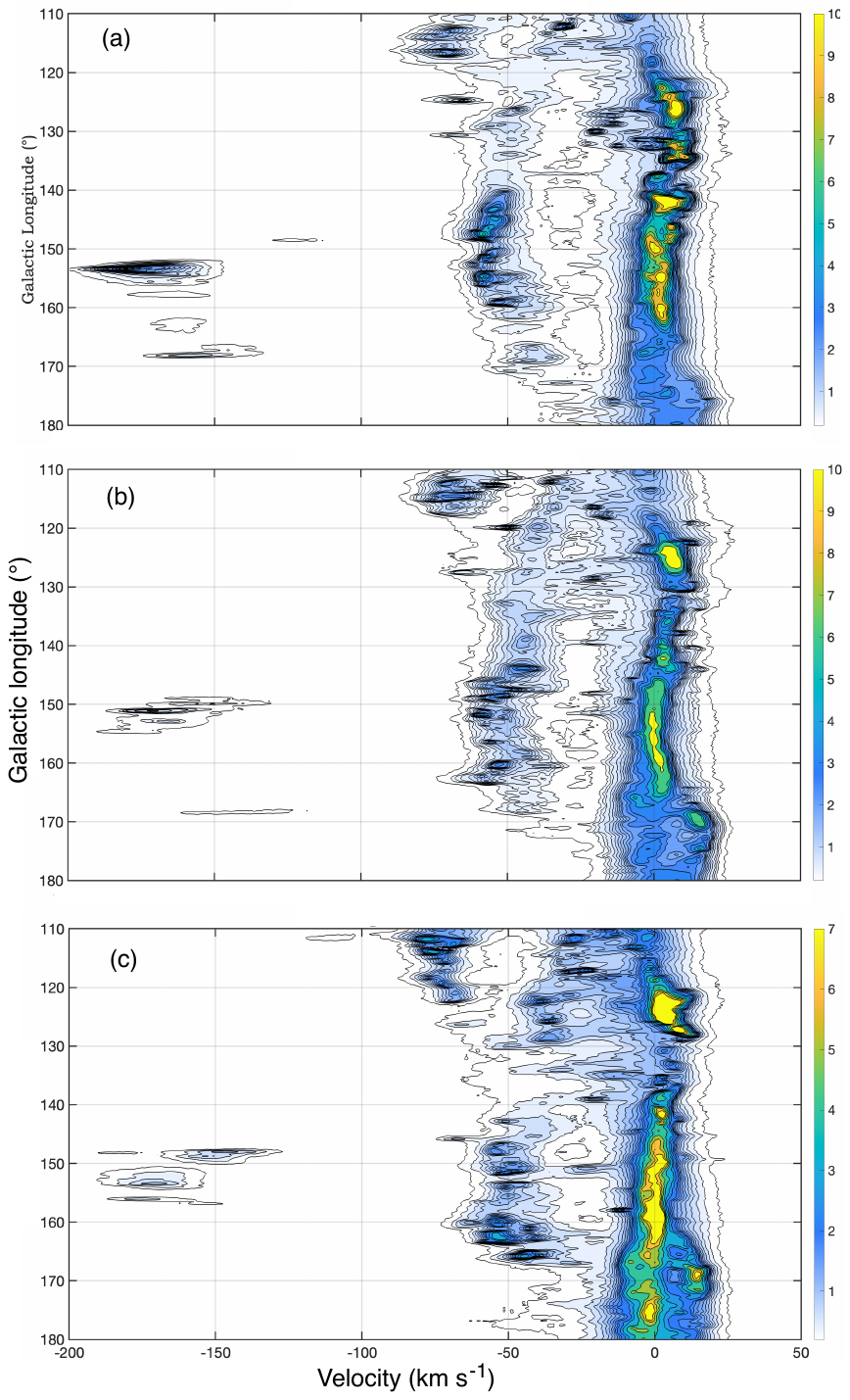}
\caption{($l$, $v$) maps at different latitudes. (a) at $b$ = 39$^{\circ}$, the curvature of the HI emission around -50 to -60 km s$^{-1}$ (the LLIV Arch) is a signature of the structure on the near side of the expanding shell. The low-velocity gas shows a clear distortion. (b) at $b$ = 37$^{\circ}$, the emission from the shell is again well-defined, bridging to low-velocity gas at around $l$ = 131 $^{\circ}$ and 170$^{\circ}$. The bright feature at $l$ = 125 $^{\circ}$, +5 km s$^{-1}$ is part of the NCP Loop. This map also shows the distortion of the low-velocity peak to positive velocities from $l$ = 120$^{\circ}$ to 145$^{\circ}$. The low-velocity gas has a wavelike velocity gradient between the location of the emission from the NCP Loop area around $l$ = 120$^{\circ}$ to the edge of the map at $l$ =180$^{\circ}$. (c) at $b$ = 35$^{\circ}$ shows the continuity of the HI around -55 km s$^{-1}$ linking with low-velocity gas at $l$ = 130$^{\circ}$ and 170$^{\circ}$. Emission from the NCP Loop is still prominent at around $l$ =125$^{\circ}$. Legends in Kelvins.
}
\end{figure}
\clearpage

\begin{figure}
\figurenum{4}
\epsscale{1.0}
\plotone{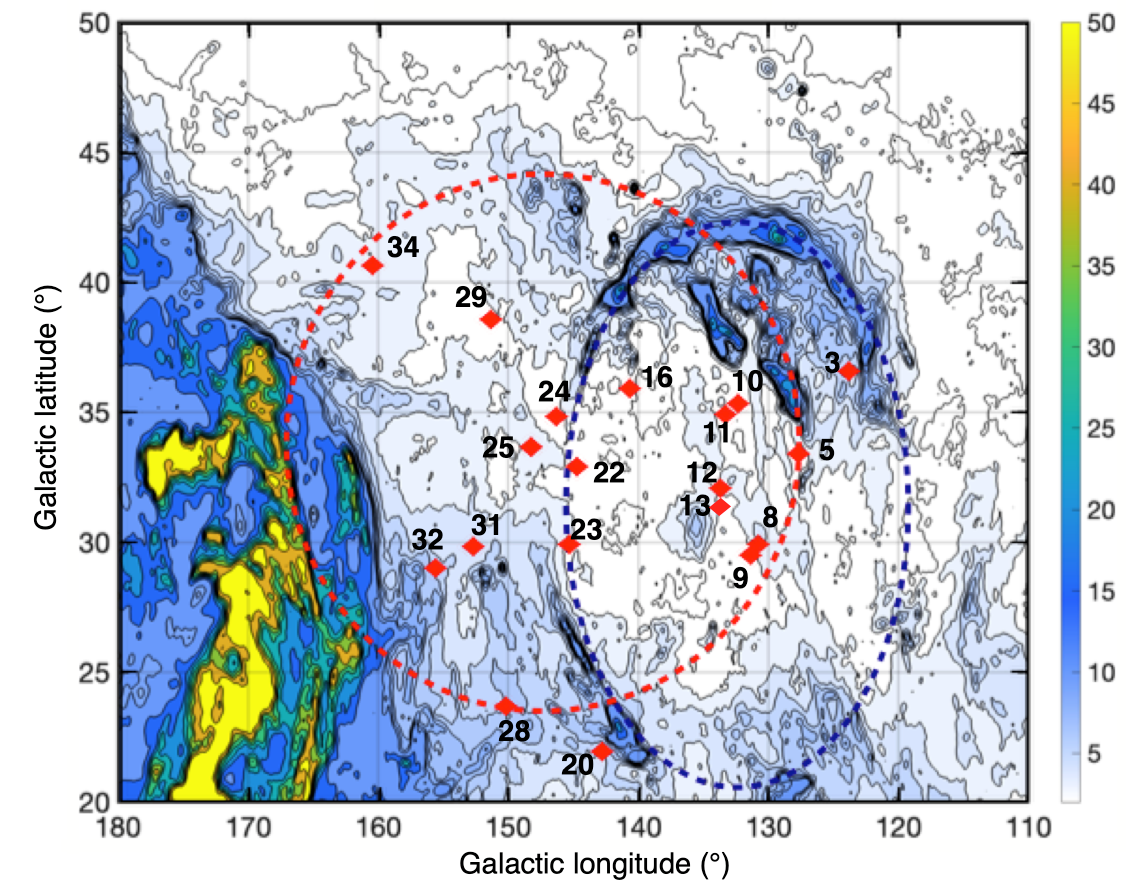}
\caption{A schematic showing the intersection of the two cavities as seen in the panel in Fig. 1k at +14 km s$^{-1}$. The red dashed circle depicts the Circular Cavity, and the blue dashed ellipse depicts the Oval Cavity. The location of the spectroscopic binary systems at distances closer than 700 pc are indicated by red filled diamonds with the adjacent numbers referring to the entries on Tables 2 - 4.  The two binary systems of greatest interest are Gaia DR3 1093757200530267520 (System \#25) close to the center of the Circular Cavity, and Gaia DR3 1144019690966028928 (System \#13) close to the center of the Oval Cavity.
}
\end{figure}

\clearpage

\begin{figure}
\figurenum{5}
\epsscale{0.8}
\plotone{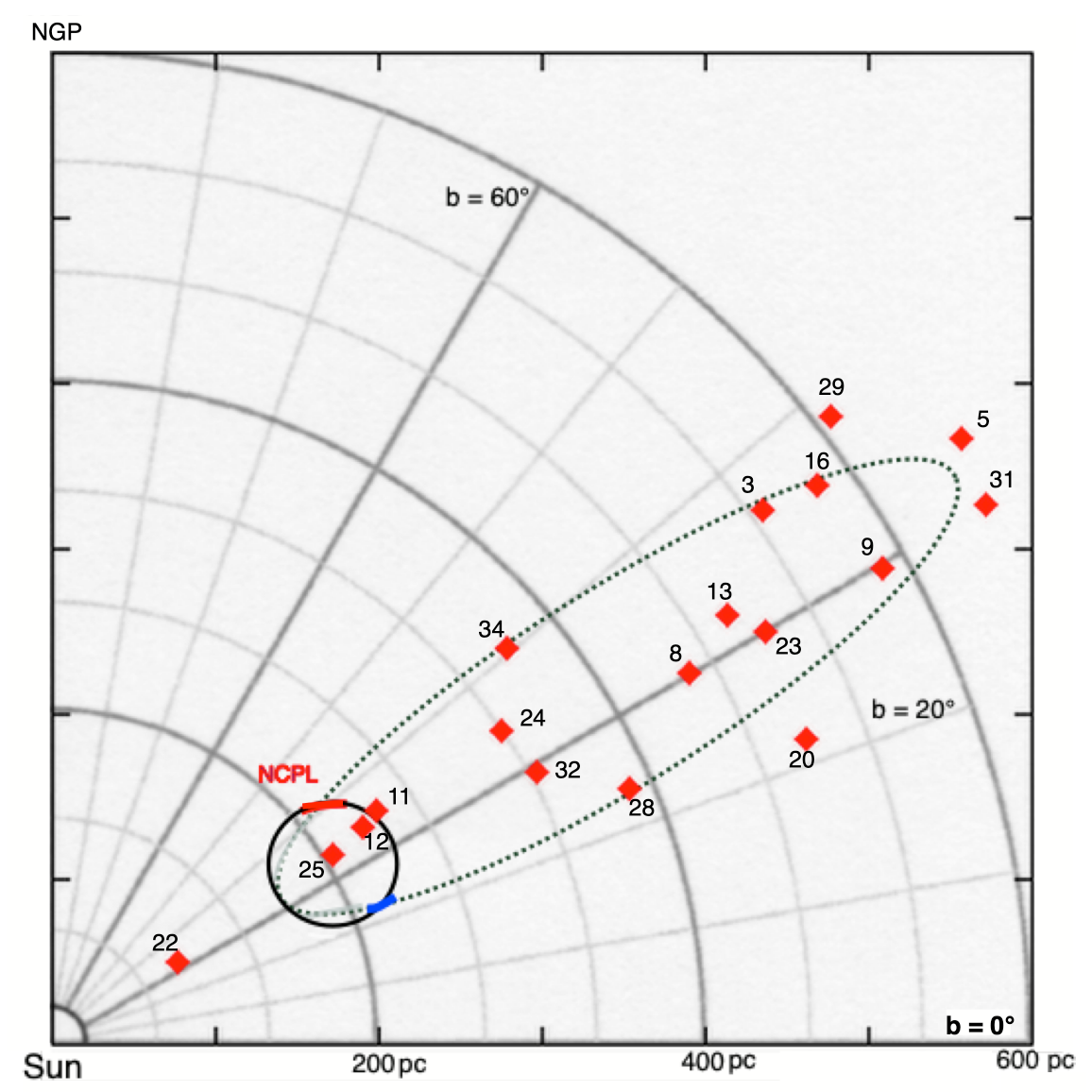}
\caption{Polar projection at around $l$ = 135$^{\circ}$ showing in a schematic of the two cavities seen in Fig. 4. The solid black circle represents the Circular Cavity and the dashed ellipse depicts the Oval Cavity with the scale based on the suggestion by \citet{Marchal2023}.  Gaia DR3 1093757200530267520 (System \#25) at a distance of 207 pc (Table 2), sets the scale for the Circular Cavity. Several binary systems with potential neutron stars in orbit (see text) lie within the outlines of the Oval Cavity, which would account for its elongation along the line-of-sight.  Also indicated are two regions where the Circular Cavity and the Oval Cavity interact - the NCP Loop in red and the lower arc in blue. 
}
\end{figure}

\end{document}